\documentclass[review,times,12pt]{elsarticle}

\usepackage{epsfig}
\usepackage{graphicx}
\usepackage{times}
\usepackage{amssymb}
\usepackage{lineno}

\journal{Physics Letters B}

\begin{document}

\begin{frontmatter}

\title{Nuclear Scaling and the EMC Effect}

\author[JLab]{D. W. Higinbotham\corref{cor1}}
\ead{doug@jlab.org}
\cortext[cor1]{Corresponding author}

\author[JLab]{J. Gomez}
\address[JLab]{Jefferson Lab, Newport News, VA 23606, USA}

\author[TelAviv]{E. Piasetzky}
\address[TelAviv]{Tel Aviv University, Tel Aviv 69978, Israel}

\begin{abstract}
Results of recent EMC effect measurements and
nuclear scaling measurements 
have both been attributed to local nuclear density effects and 
not properties of the bulk nuclear system.
This lead us to the phenomenological observation that 
the ratio of the slopes in the $0.3 < x_B < 0.7$ EMC data
scale as the ratio of the $x_B > 1$ nuclear scaling plateaus.  
Using this correlation,
we developed a
phenomenological relation
which reproduces the general trends and features of the EMC effect for
nuclei from $^3$He to $^{56}$Fe.     
\end{abstract}

\end{frontmatter}

\section*{Introduction}

In 1983, the European Muon Collaboration measured the deep-inelastic per nucleon 
cross section ratios of heavy nuclei to deuterium over a broad kinematic
range~\cite{Aubert:1983xm}.  
This ratio revealed an unexpected structure which was 
subsequently confirmed by SLAC~\cite{Gomez:1993ri} and became known as the EMC effect.
These cross section ratios are typically plotted as a 
function of the Bjorken scaling variable $x_B$ where $x_B = {\rm{Q}}^2/2m\omega$ 
with Q$^2$ the four momentum transferred to the system,  $\omega$ the energy 
transfer, and $m$ the mass of a proton.   Plotted this way, there is 
a sharp rise for $x_B > 0.8$, a persistent dip around $x_B \sim 0.75$ 
and a gentle rise from $x_B$ from 0.7 to 0.3.
These results have generated considerable experimental and theoretical interest,
with the explanation for the effect generally
attributed to a change in quark distributions in nuclei or 
persistent nuclear effects~\cite{Geesaman:1995yd,Norton:2003cb}.

New high-precision EMC effect experimental data on light nuclei 
from Jefferson Lab~\cite{Seely:2009gt} suggest that the slope in
the $0.3 < x_B < 0.7$ region is a local density effect and not a bulk property
of the nuclear medium. 
At similar four-momentum transfers, data from SLAC and Hall B have shown that the $x_B > 1$ nuclear scaling plateaus,
which are related to the high-momentum components of the nuclear wave function~\cite{Bosted:1982gd},
are also a local density effect.
With these two observations in mind, we noticed that
that the ratio of the slopes in the 
EMC $0.3 < x_B < 0.7$ data scale as the ratio of the $x_B > 1$
nuclear scaling plateaus.   

Theory calculations have been done that use nucleon momentum distributions
to try to reproduce the EMC effect~\cite{Kumano:1989eh,CiofidegliAtti:1990dh,Gross:1991pi,Rinat:2004ia},
The work of Kumano and Close even showed that local nuclear density effects and quark rescaling
models are compatible, but that more data was needed to see if a true correlation exists~\cite{Kumano:1989eh}.
In this work,
we have used the new nuclear scaling data as a phenomenological tool for determining
the magnitude of the high-momentum part of
the nucleon momentum distribution for various nuclei.\footnote{By the Heisenberg uncertainty principle,
$\Delta x \Delta p \ge \hbar/2$, the 
wide distribution of the high-momentum tail, $\Delta p$,
implies a small local density, $\Delta x$, as compared
to the narrower mean field momentum distribution.}
We also make use of the observed dominance of initial-state quasi-deuteron correlations
in nuclei~\cite{Piasetzky:2006ai,Subedi:2008zz} to make a kinematic connection between the
$x_B > 1$ scaling plateaus and the EMC $0.3 < x_B < 0.7$ slopes.

\section*{Nuclear Scaling}

In electron scattering, with $x_B > 1$ and Q$^2 > 1~[GeV/c]^2$, 
Q$^2$ independent plateaus have been observed in inclusive per-nucleon cross-section ratios.
This inclusive nuclear scaling was first observed at SLAC~\cite{Sick:1980ey,Day:1987az,Frankfurt:1993sp} 
and subsequently by experiments at Jefferson Lab~\cite{Egiyan:2003vg,Fomin:2008iq}.  
In general, this scaling is attributed to initial-state nucleon-nucleon correlations
causing the high-momentum tail in all nuclei.
Assuming this is correct, then the magnitude of the nuclear scaling ratios give the 
relative strength of the nucleon-nucleon correlations in the 
various nuclei~\cite{Sargsian:2002wc,Frankfurt:2008zv}. 

Confirming this hypothesis,
high-momentum recoiling nucleons have been observed in $^{12}$C two-nucleon knock-out
experiments both with hadronic~\cite{Piasetzky:2006ai} and leptonic 
probes~\cite{Subedi:2008zz,Shneor:2007tu}.  The dominance 
of proton-neutron pairs to proton-proton pairs in these reactions 
has been attributed to initial-state tensor correlations
as calculated using realistic nucleon-nucleon potentials~\cite{Sargsian:2005ru,Schiavilla:2006xx,Alvioli:2007zz}.

The nuclear scaling plateaus in the $x_B > 1$ (e,e') ratios have been experimentally shown to start
when the magnitude of the minimum missing momentum, $p_{min}$, of the D(e,e')pn reaction
is greater than the Fermi momentum of $\sim$250 MeV/c where $p_{min}$ is the minimum 
initial momentum that the struck nucleon must have
in order to produce an $x_B > 1$ final state.  Remember, for scattering from a free nucleon, 
there can be no $x_B > 1$ values.
To calculate $p_{min}$ one uses
conservation of energy and momentum for quasi-elastic 
scattering from the deuteron $(q^{\mu}+p_d^{\mu} - p_r^{\mu})^2 = m^2$
where $q^{\mu}$  $p_d^{\mu}$ and $p_r^{\mu}$ are four momenta of the virtual photon,
deuteron and recoil nucleon, respectively, along with the constraint that 
$x_B$ = Q$^2/2m\omega$~\cite{Sargsian:2002wc,Egiyan:2003vg,Frankfurt:2008zv}.

\begin{figure}
\centering
\includegraphics[width=\linewidth]{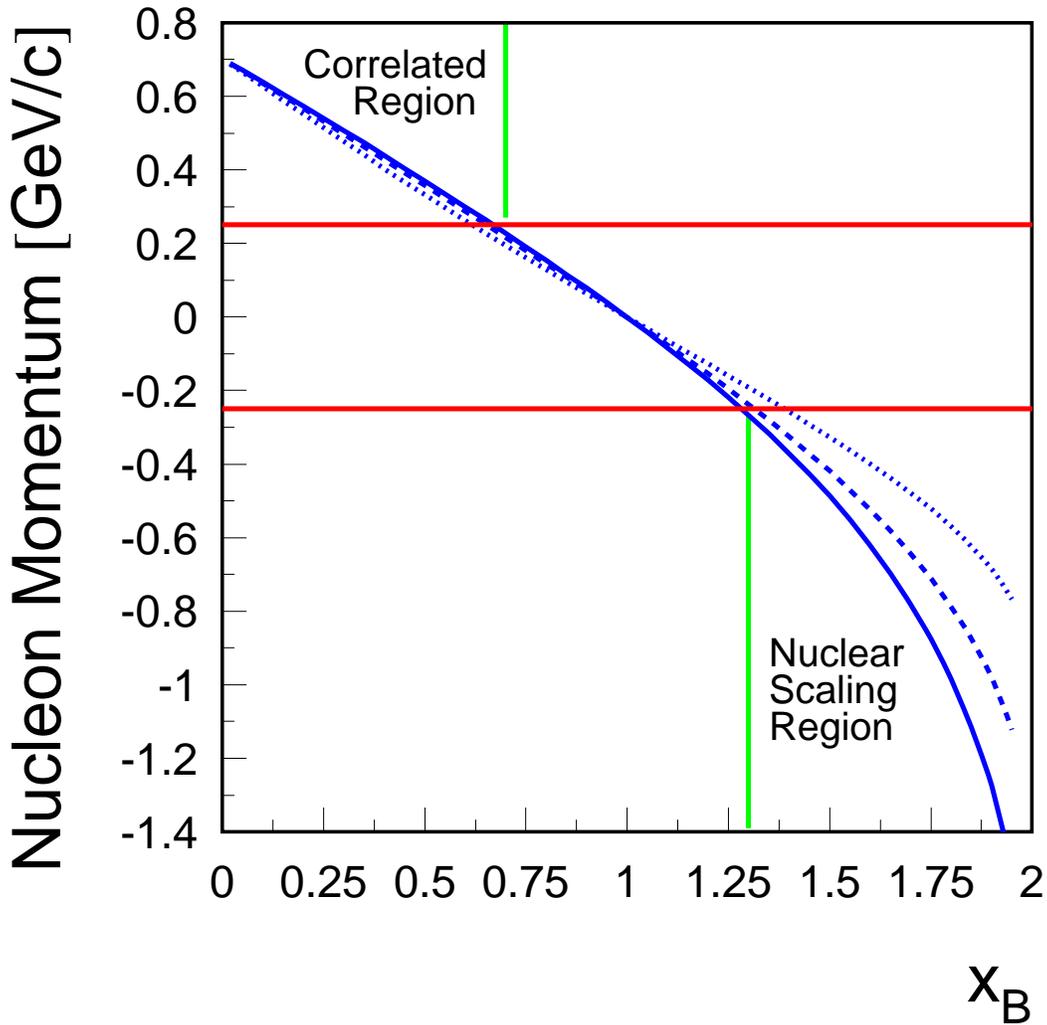}
\caption{Phase space plot of allowed initial-state nucleon momenta for the D(e,e')pn reaction 
as a function of $x_B$ assuming the initial-state is a correlated pair of moving nucleons.  
The minimum possible momentum of the nucleons, $p_{min}$,  
negative for $x_B > 1$ and positive for $x_B < 1$,
for Q$^2$ values of 4, 8, and 10 GeV$^2$ 
is shown as dotted, dashed and solid lines, respectively.  
The region between the red lines indicates the ~250 MeV/c region of simple Fermi motion.  
Beyond the Fermi region for $x_B > 1$ is where inclusive nuclear scaling has been observed. 
The correlated region is simply the kinematics of the nuclear scaling region's partner nucleons, i.e. for any
given negative nucleon momentum there must be a corresponding nucleon with a positive momentum
so that the total initial-state momentum is zero.
}
\label{emc-pmin}
\end{figure}

The phase space mapped out by this function  for various Q$^2$ is shown in Fig.~\ref{emc-pmin} where
the allowed nucleon momenta are negative for $x_B > 1$ and positive for $x_B < 1$.   
The curve gives the Q$^2$ dependent minimum kinematically allowed missing momentum, $p_{min}$, 
for the D(e,e')pn reaction as given value of $x_B$.
As the initial-state involves a correlated pair of nucleons, there is a 
corresponding $x_B < 1$ value for every $x_B > 1$ value.  
For example, for a Q$^2$ = 10 [GeV/c]$^2$ the -0.25~GeV/c point at $x_B = 1.3$ is correlated
with the 0.25~GeV/c point at $x_B = 0.7$.
The areas of correspondence between the $x_B > 1$ scaling region and the correlated $x_B < 1$ region
are labeled on Fig.~\ref{emc-pmin} as the nuclear scaling and correlated regions, respectively.

The presence of flat, Q$^2$ independent plateaus in the $x_B > 1$ A(e,e')/D(e,e') nuclear scaling 
data simply indicates that the underlying functional form of the high-momentum distribution for
various nuclei is only different by a scale factor.  
Thus, to a good approximation, one can use nuclear scaling results along with
a realistic proton-neutron pair momentum distribution 
to calculate the absolute strength of the high-momentum tail for various nuclei.
To do this, the Argonne v-18 potential~\cite{Wiringa:1994wb} 
was used to calculate the deuteron's nucleon momentum distribution, $n_D(p)$,
where the function was normalized such 
that $1 = 4\pi \int_{0}^{\infty} p^2 \times n_D(p) \times dp$
and the functional form 
of the high-momentum distribution, $n_A(p)$, of nuclei A was approximated for $p > 250$~MeV/c by
\begin{equation}
\label{eq-momentum}
n_A(p) = n_D(p) \times C^A
\end{equation}
where $C^A$ is the Q$^2$ independent $x_B > 1$ per-nucleon nuclear scaling 
ratio~\cite{CiofidegliAtti:1990dh} which can be determined directly from
experimental data~\cite{Frankfurt:1993sp,Egiyan:2003vg}. 
Thus, Eq.~\ref{eq-momentum} allows a phenomenological calculation of the high-momentum distribution
for any nucleus for which the nuclear scaling ratio has been measured.
Since most of the new nuclear scaling data is taken as a  ratio to $^3$He instead of D, 
we did need to use the calculated value of 2 to go from $^3$He to D~\cite{Egiyan:2003vg,Fomin:2008iq}; 
and have used nominal values of 2, 4, 4.8, 5.7 for the
$^3$He, $^4$He, $^{12}$C, and $^{56}$Fe to deuterium scaling ratios, $C^A$, respectively.

\section*{The EMC Effect}

For electron scattering on a free nucleon, such as the proton, 
the observed (e,e') reaction can kinematically cover a range from $0 < x_B < 1$;  
but for nuclei with an atomic mass A, the range can go from $0 < x_B < A$.
As a thought experiment, one can consider a one-dimensional quasi-deuteron
initial state made up of nucleons moving with
relative momentum of $\pm250$~MeV/c.
When probed in quasi-elastic kinematics at a Q$^2$ of 10~[GeV/c]$^2$, 
this initial state would yield two peaks, one at $x_B$ = 0.7
and another at $x_B$=1.3.
When that same initial-state is probed in deep-inelastic kinematics, it would produce
two super-imposed  spectrum, one starting at $x_B$ = 1.3 (i.e. the maximum $x_B$ of one of
the nucleons) and a second starting at $x_B$ = 0.7.
Thus, in this thought experiment, only one nucleon of the two can contribute to an observed cross section
in the $x_B > 0.7$ region while both contribute to the $x_B < 0.7$ region.

By using the experimental result that high-momentum components of the initial-state
are dominated by quasi-deuteron nucleon pairs~\cite{Subedi:2008zz}, 
the correspondence between  initial-state momenta and values of $x_B$ 
as shown in Fig.~\ref{emc-pmin} for a deuteron can be used to make a general connection
between the nuclear scaling region and the $0.3 < x_B < 0.7$ region.
To do this, we simply assume that initial-state
nucleons either contribute to the cross section or are in a  
high-momentum state, like in the thought experiment, that kinematically cannot 
contribute.  
Thus, for the $0.3 < x_B < 0.7$ region the per-nucleon cross section ratio
can be written as
\begin{equation}
\frac{\sigma_A}{\sigma_D} = \frac{1 - \mathcal{P}_A(x_B)}{1 - \mathcal{P}_D(x_B)}
\label{final-answer}
\end{equation} 
where $\mathcal{P}_A(x_B)$ is the $x_B$ dependent fraction of
the inital-state momentum distribution, $n_A(p)$, that is kinematically 
forbidden to contribution to the cross section.
To calculate the $\mathcal{P}_A(x_B)$,
we use the nucleon momentum distribution, $n_A(p)$, from Eq.~\ref{eq-momentum}
and integrate for a given value of $x_B$ from $p_{min}$ to infinity
\begin{equation}
\mathcal{P}_A(x_B) = 2\pi \int^{\infty}_{p_{min}} p^2 \times n_A(p) \times dp
\end{equation}
where the function has been divide by two to count only the positive $p_{min}$ contribution.  
The results of the integration for a Q$^2$ = 10 [GeV/c]$^2$ for 
various nuclei and values of $x_B$ are shown in Table~\ref{lowxtable}.
For the $x_B > 1$ nuclear scaling plateau region we use
\begin{equation}
\frac{\sigma_A}{\sigma_D} = C^A
\label{eq:high}
\end{equation}
where the values of C$^A$ are obtained from experimental data.

\begin{table}
\centering
\begin{tabular}{lcccccc}
$x_B$ & $|p_{min}|$ & Deuteron & $^3$He& $^4$He &$^{12}$C & $^{56}$Fe \\ 
      &  [MeV/c]   & [\%]     & [\%]  & [\%]   &  [\%]   & [\%]      \\ \hline
0.70  &  250       & 2.5      & 5.0   &  10    &  12     & 14.3      \\
0.68  &  275       & 2.0      & 4.0   &  8.0   &  9.6    & 11.4      \\
0.6   &  300       & 1.6      & 3.2   &  6.4   &  7.7    &  9.1      \\
0.5   &  370       & 1.25     & 2.5   &  5.0   &  6.0    &  7.3      \\ 
0.45  &  400       & l.1      & 2.2   &  4.4   &  5.3    &  6.5      \\
0.4   &  440       & 0.9      & 1.8   &  3.6   &  4.3    &  5.2      \\
0.3   &  500       & 0.65     & 1.3   &  2.6   &  3.1    &  3.7      \\
\end{tabular}
\caption{The percent of $\mathcal{P}_A(x_B)$, the initial-state momentums fraction above $p_{min}$,
for values of $x_B$ for the D(e,e')pn reaction and for Q$^2$ = 10~[GeV/c]$^2$.
The Argonne v-18 nucleon-nucleon potential was used to generate the momentum distribution for the deuteron.
For the other nuclei, the relation $n(k)_A = n(k)_d \times C^A$ is used where $C^A$ is 
the taken from $x_B > 1$ inclusive nuclear scaling ratios
of $^3$He, $^4$He, $^{12}$C, and $^{56}$Fe to deuterium of 2, 4, 4.8 and 5.7, 
respectively~\cite{Frankfurt:1993sp,Egiyan:2003vg}.}
\label{lowxtable}
\end{table}

\begin{figure}
\centering
\includegraphics[width=\linewidth]{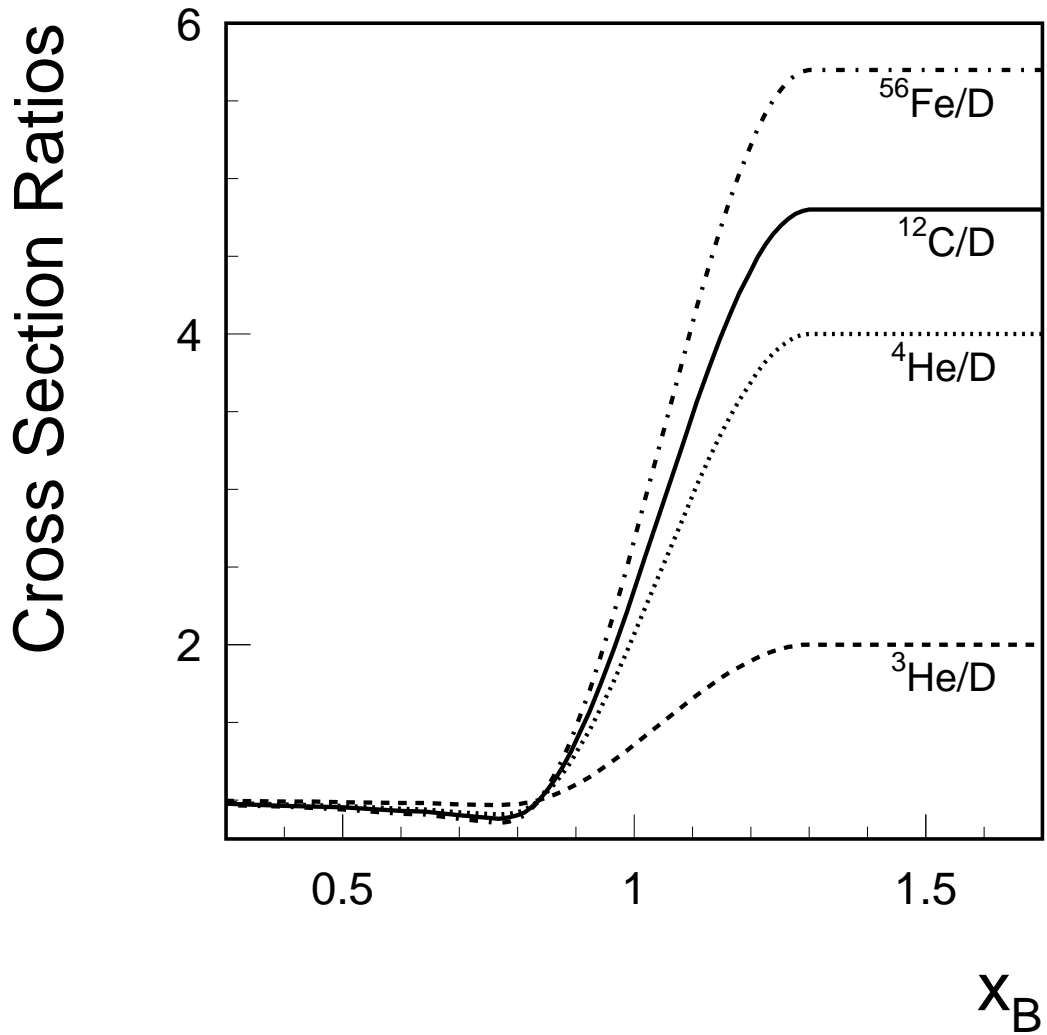}
\caption{Plot of the expected cross section ratio function for a $Q^2$ of 10 [GeV/c]$^2$ from $0.3 < x_B < 1.7$ 
for $^{56}$Fe/D (dashed-dotted curve), $^{12}$C/D (solid curve), $^4$He/D (dotted curve), 
and $^3$He/D (dashed curve) where the $0.3 < x_B < 0.7$ slopes have been calculated
using Eq.~\ref{final-answer} and the $0.7 < x_B > 1.3$ region has been determined by interpolation
to the $x_B > 1.3$ nuclear scaling plateaus.}
\label{emc-effect}
\end{figure}

Figure~\ref{emc-effect} shows the result of Eq.~\ref{final-answer} and Eq.~\ref{eq:high}
for a Q$^2$ of 10~[GeV/c]$^2$ for various nuclei.   
In order to have the function cover the entire range from $0.3 < x_B < 1.7$,
it has been assumed in the limit of high Q$^2$ that the simple Fermi motion can be ignored
and a smooth interpolation can be made between the $x_B = 0.7$ and $x_B = 1.3$ nuclear
scaling plateau,
with the only constraint to the function being that the value at $x_B = 0.8$ equal the value at $x_B = 0.7$
to cause the known inflection around $x_B = 0.75$.
The idea of simply connecting the regions with a smooth function is motivated by data
where the typical dip around $x_B = 1$ in the quasi-elastic (e,e') ratios 
becomes less pronounced as Q$^2$ increases (see Fig. 5.17 of~\cite{Fomin:2008iq}).
In fact, the BCDMS collaboration's Q$^2$ = 50~[GeV/c]$^2$ carbon data
follows a smooth exponential in the $x_B \approx 1$ region~\cite{Benvenuti:1994bb}.

\begin{figure}
\centering
\begin{tabular}{lr}
\includegraphics[width=0.45\linewidth]{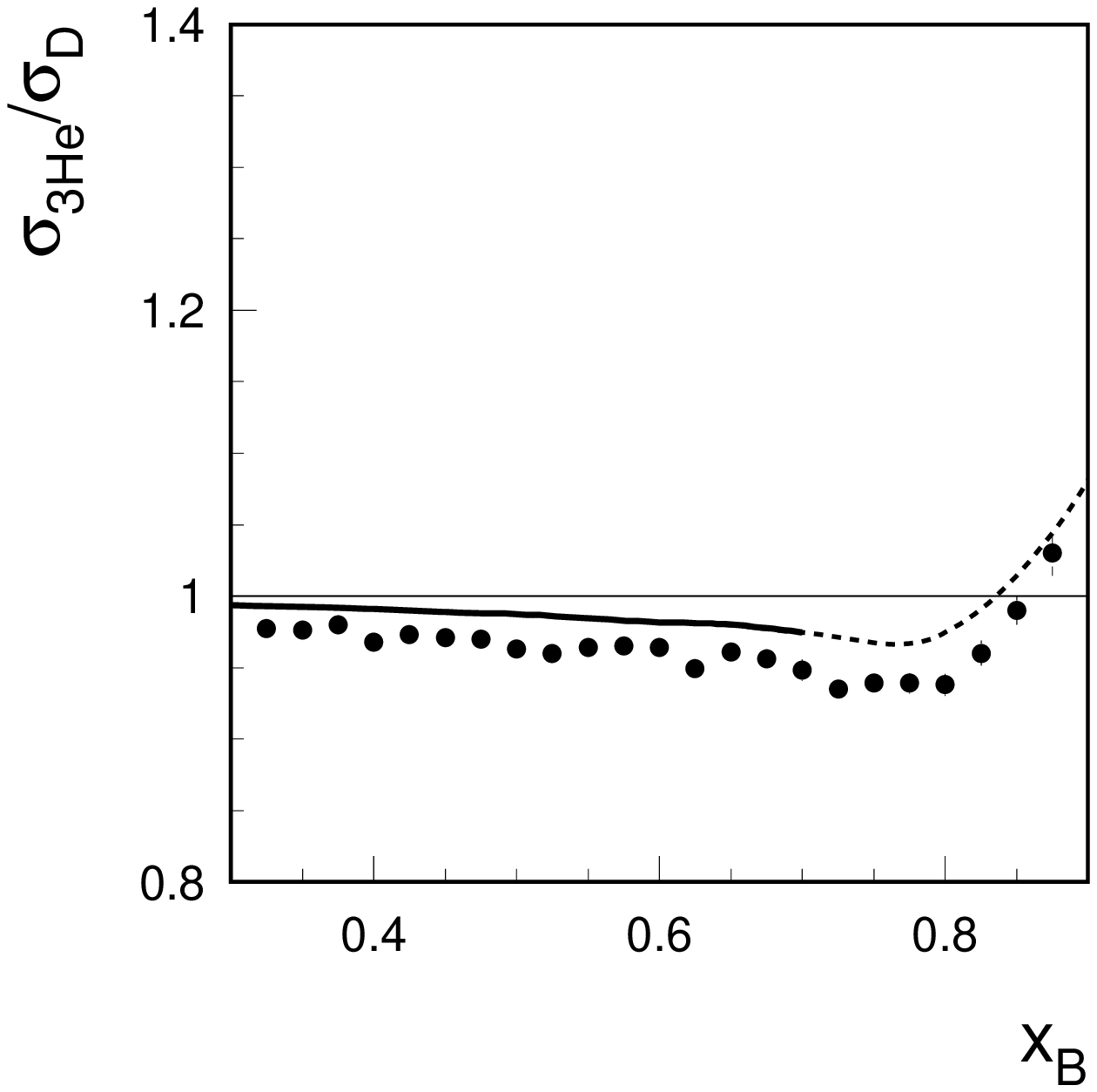} &
\includegraphics[width=0.45\linewidth]{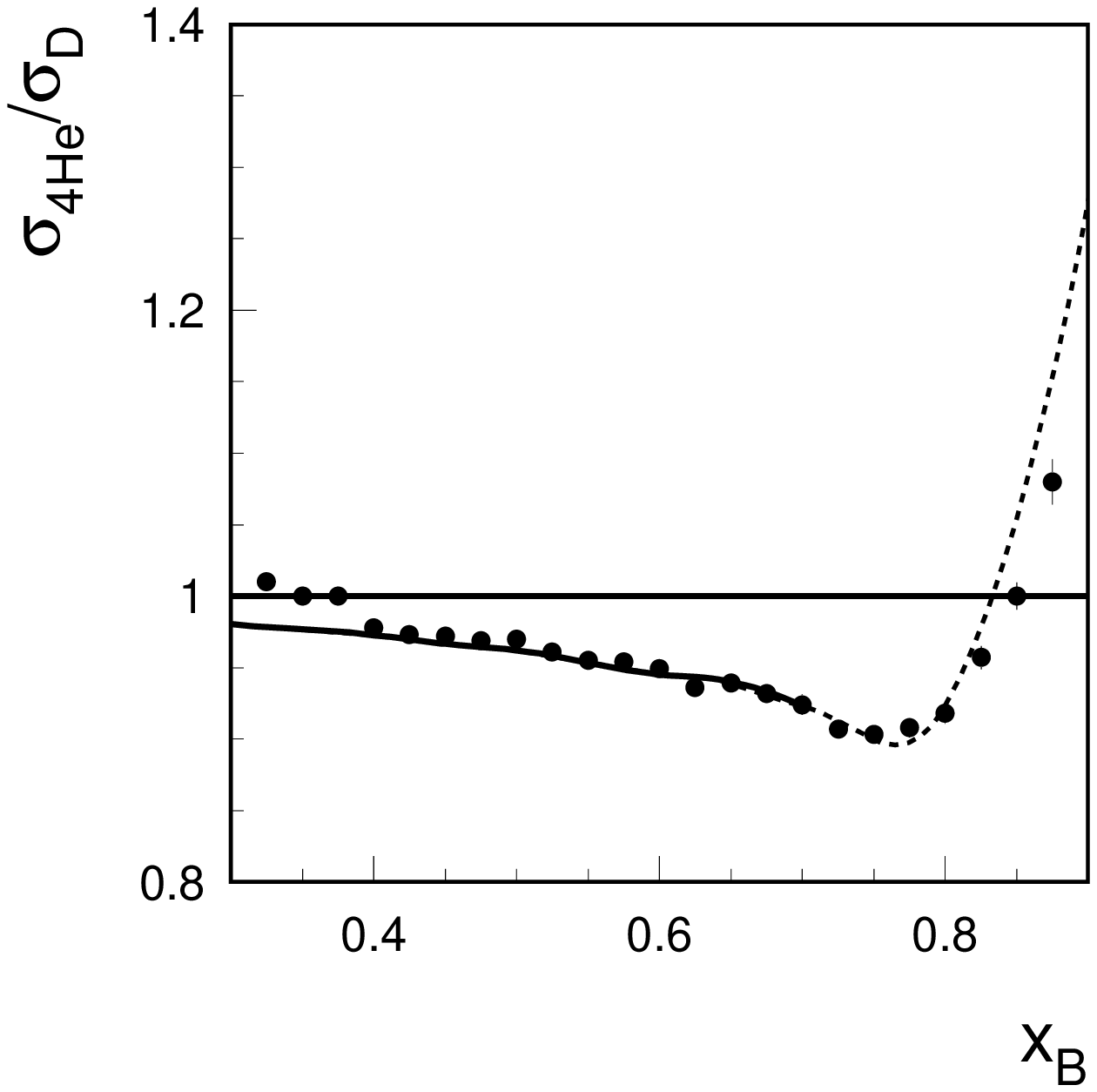} \\
\includegraphics[width=0.45\linewidth]{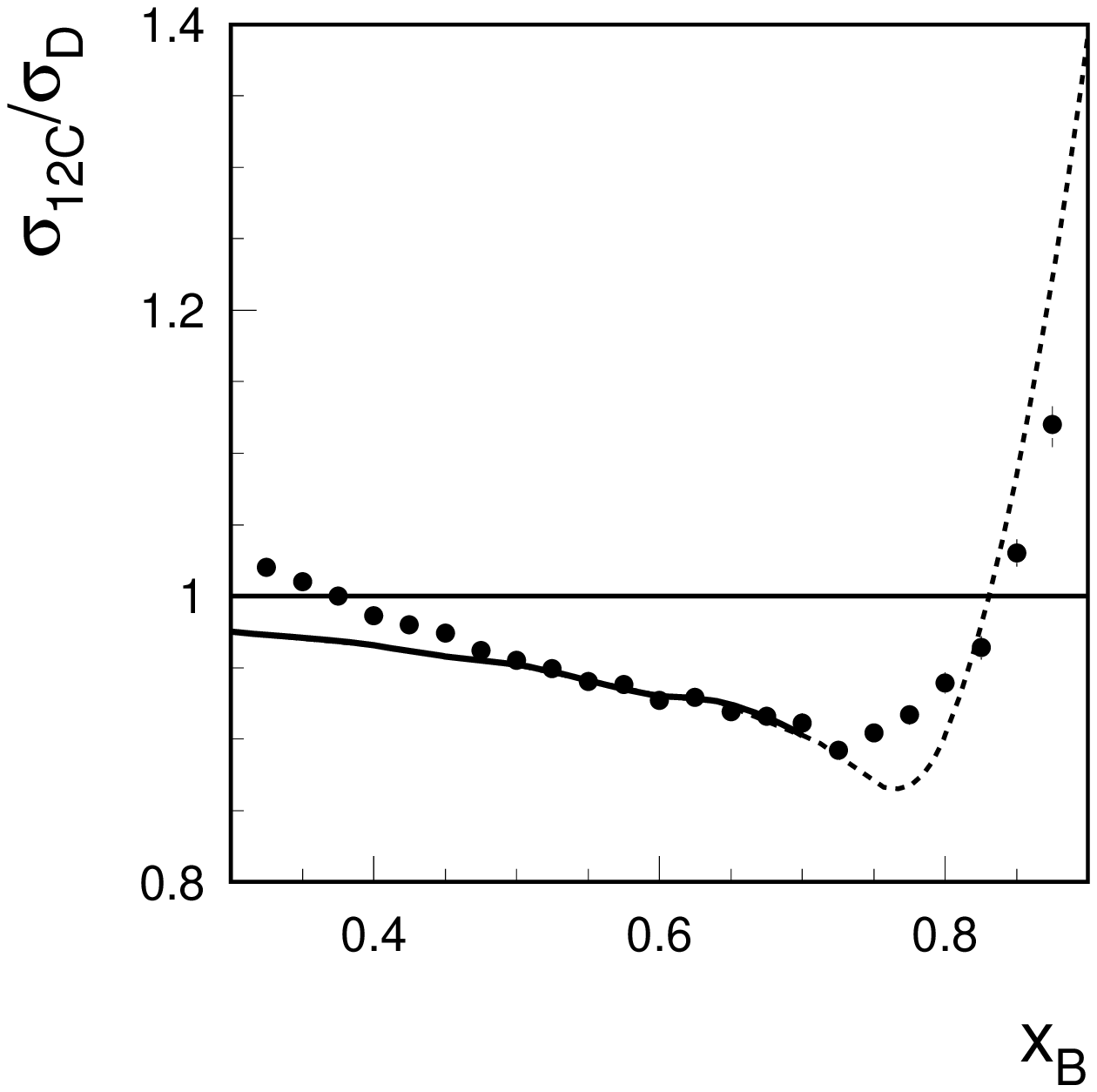} &
\includegraphics[width=0.45\linewidth]{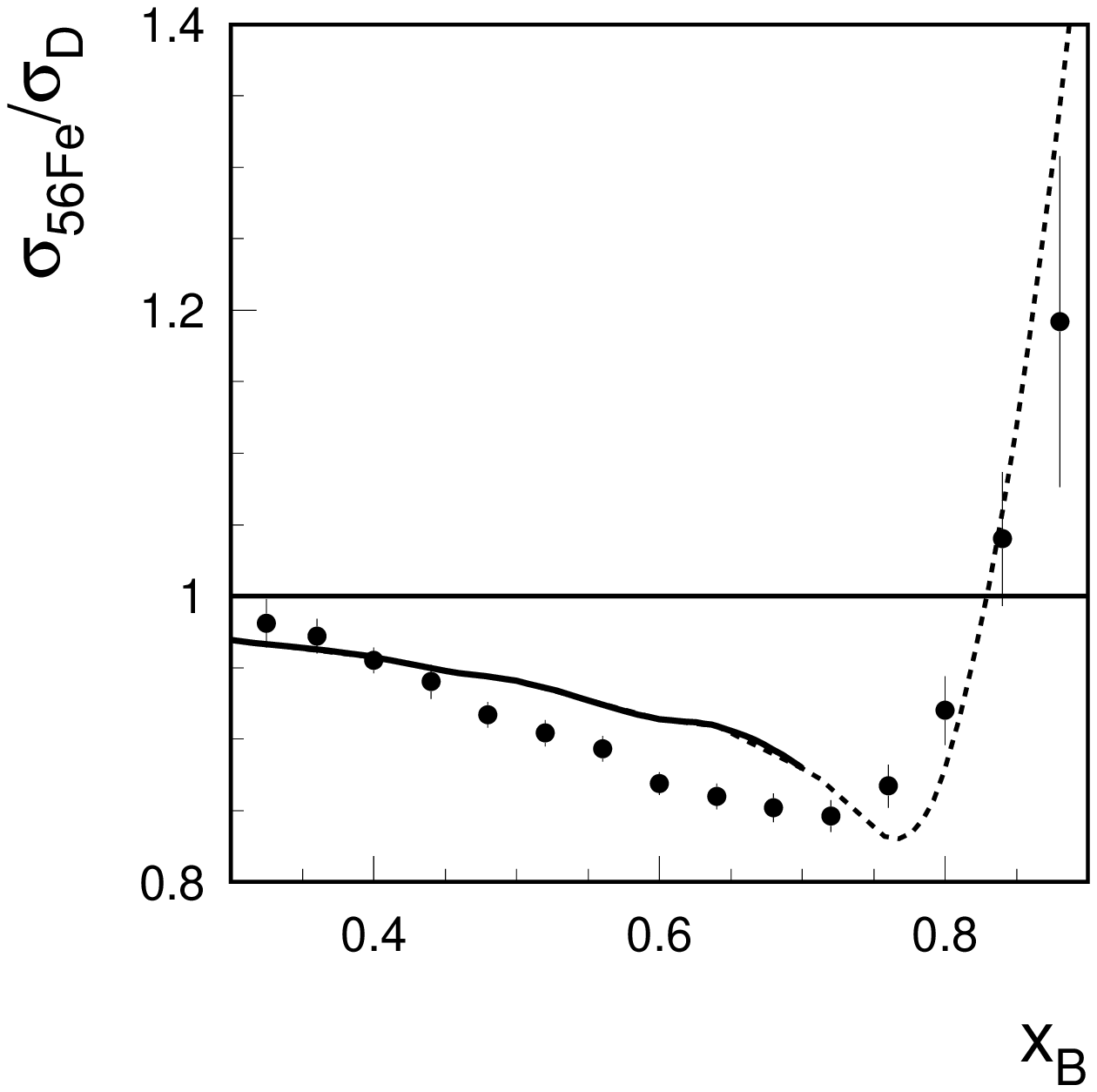} 
\end{tabular}
\caption{The solid line is the calculation of the ratio while the dashed is a simple
interpolation between the $x_B < 1$ and $x_B > 1$ regions.
The $^3$He, $^4$He, and $^{12}$C data are from Jefferson Lab~\cite{Seely:2009gt} and
the $^{56}$Fe data from SLAC~\cite{Gomez:1993ri}.  The calculation gets the general
shape of the downward slope in the $0.3 < x_B < 0.7$ region and the interpolation 
to the nuclear scaling plateau reproduces
the theoretically difficult to reproduce upward slope of the EMC effect in the $x_B > 0.8$ region. }
\label{idea-all}
\end{figure}

Figure~\ref{idea-all} shows EMC data along with result of the $0.3 < x_B < 0.7$ calculation from Eq.~\ref{final-answer},
along with the interpolation to the $x_B > 1$ plateaus.  The plots are all made with the same scales and it can be seen
that the general trends in the data are matched by this simple phenomenological model.

\section*{Summary}

In summary, we note that the ratio of nuclear scaling plateaus and the ratio
of $0.3 < x_B < 0.7$ data following the same pattern.  
Since recent leptonic and hadronic data which has shown that 
the nuclear scaling region is dominated by high-momentum initial-state quasi-deuteron pairs, 
we tried using the kinematics of pairs, to see if that same initial-state could be produce
an EMC like effect by assuming the offshell state would not contribute to the measured cross section.
Interestingly, this simple model
reproduces the general features of the EMC effect, with a slope that in the $0.3 < x_B < 0.7$ 
region decreasing and the upward interpolation for $x_B > 0.8$ generally agreeing with data.
This phenomenological result seems to strengthen the hypothesis that both
the EMC effect and nuclear scaling are due to local density effects and related to the 
high momentum components of the nuclear wave function.
\section*{Acknowledgements}

This paper was inspired by a conversation between D.H. and Hugh Montgomery where
the EMC effect and nuclear scaling were discussed and plots of $x_B <1$ and 
$x_B >1$ data were shown taped together.
The paper is dedicated to our deceased colleague Kim Egiyan who pushed us to do the inclusive
hadronic scaling experiments at Jefferson Lab.
We thank Misak Sargsian for providing 
the deuteron momentum distribution function and for the constructive
comments from Franz Gross and Jerry Miller.
This work was supported by the U.S.\ Department of Energy,
the U.S.\ National Science Foundation, the Israel Science Foundation, 
and the US-Israeli Bi-National Scientific
Foundation.  Jefferson Science Associates operates
the Thomas Jefferson National Accelerator Facility under DOE
contract DE-AC05-06OR23177.



\end{document}